\documentclass[twocolumn,showpacs,preprintnumbers,amsmath,amssymb,final]{revtex4}

\usepackage{graphicx}% Include figure files
\usepackage{dcolumn}% Align table columns on decimal point
\usepackage{bm}% bold math

\def\BEq{\begin{equation}}
\def\EEq{\end{equation}}
\def\BEqA{\begin{eqnarray}}
\def\EEqA{\end{eqnarray}}
\def\BEn{\begin{enumerate}}
\def\EEn{\end{enumerate}}
\def\BWT{\begin{widetext}}
\def\EWT{\end{widetext}}
\def\a{\alpha}

\begin{document}

%\preprint{APS/123-QED}

\title{
Quantum computation with prethreshold superconducting qubits: 
Single-excitation subspace approach
}

\author{Andrei Galiautdinov$^{1}$ and Michael R. Geller$^{2}$}

\affiliation{
$^{1}$Department of Electrical Engineering,
University of California, Riverside, California
92521, USA
\\
$^{2}$Department of Physics and Astronomy,
University of Georgia, Athens, Georgia
30602, USA
}

\date{\today}% It is always \today, today,

\begin{abstract}

We describe an alternative approach to quantum computation that 
is ideally suited for today's sub-threshold-fidelity qubits, and 
which can be applied to a family of hardware models that includes 
superconducting qubits with tunable coupling. In this approach, the 
computation on an $n$-qubit processor is carried out in the 
$n$-dimensional single-excitation subspace (SES) of the full  
$2^n$-dimensional Hilbert space. Because any real Hamiltonian 
can be directly generated in the SES 
[E.~J.~Pritchett  {\it et al.}, arXiv:1008.0701], high-dimensional 
unitary operations can be carried out in a single step, bypassing 
the need to decompose into single- and two-qubit gates. Although 
technically nonscalable and unsuitable for applications (including Shor's) 
requiring enormous Hilbert spaces, this approach would make practical 
a first-generation quantum computer capable of achieving significant 
quantum speedup.

\end{abstract}

\pacs{03.67.Lx, 85.25.-j}    

\maketitle

%\tableofcontents

\section{Introduction}

In the standard gate-based model of quantum computation with $n$ qubits, 
the initial state of the (closed) system is 
represented by a $2^n$-dimensional complex vector $\psi$, and the 
computation is described by a unitary time-evolution operator $U$. 
Running the quantum computer implements the map
\begin{equation}
\psi \rightarrow U \psi.
\end{equation}
A key feature of this approach to quantum information processing is the 
exponential classical information 
storage capacity of the 
wave function $\psi$. However, the number of one- and two-qubit gates 
required to implement an arbitrary element 
of the unitary group ${\rm U}(2^n)$ is at least $2^{2n}$, making the 
construction of an arbitrary $U$ inefficient  (using 
elementary gates). The goal of quantum algorithm design is to compute 
interesting cases of $U$ with a polynomial 
number of elementary gates; two important examples are the algorithms 
for factoring \cite{ShorSIAMJC97} and quantum 
simulation \cite{LloydSci96}.
Although these algorithms are efficient, the number of gates required 
for interesting applications is still large
and error-corrected qubits are therefore required \cite{Neilsen2000}.

Central to this gate (or quantum circuit) model of quantum computation 
is the idea that a set of elementary one- and 
two-qubit gates can be used to construct an arbitrary high-dimensional 
$U$. But why should we use such a 
decomposition in the first place? The reason is that the Hamiltonians 
nature usually provides for us have one- and two-qubit terms
(one-body and two-body operators). Consider, for example, the following 
somewhat generic solid-state
model of an array of $n$ coupled  qubits,
\begin{eqnarray}
H_{\rm qc} = \sum_{i} \epsilon_i c_i^\dagger c_i
+\sum_{i < j}g_{ij} \, J_{\mu\nu} \, \sigma^\mu_i\otimes\sigma^\nu_j,
\label{tcsq model}
\end{eqnarray}
written in the basis of uncoupled-qubit eigenstates \cite{notationNote}. 
Here $ i,j = 1, 2, \dots, n,$ and  $\mu, \nu = x, y, z.$
The $\epsilon_i$ are the uncoupled qubit energies, $g_{ij}$ (with $i \neq j$) 
are qubit-qubit interaction strengths, and $J_{\mu\nu}$ is a fixed dimensionless 
tensor determined by the hardware. 

It is clear from   
(\ref{tcsq model}) that one- and two-qubit operations are naturally generated 
in this system, and we know 
(from universality \cite{LloydPRL95,BarencoPRA95}) that this is sufficient. 
If the hardware also provided controllable terms of the form 
\begin{eqnarray}
\sum_{i jk }g_{ijk} \, J_{\mu\nu\lambda}\sigma^\mu_i \otimes \sigma^\nu_j 
\otimes \sigma^\lambda_k  ,
\label{three-qubit interactionl}
\end{eqnarray}
then three-qubit gates could be used as primitives as well.
 
 \section{QUANTUM COMPUTATION IN THE SES}
 
The idea we explore here is to perform a quantum computation in the 
$n$-dimensional single-excitation 
subspace (SES) of the full $2^n$-dimensional Hilbert space. This is the 
subspace spanned by the computational
basis states 
\begin{equation}
|m) \equiv c_m^\dagger |00 \cdots 0\rangle = |0 \cdots 1_m  \cdots 0\rangle, 
\end{equation}
with $m=1,2,\dots,n.$ Experimentally, it is possible to prepare the quantum 
computer in the SES, and it will
remain there with high probability as long as (i) the coupling strengths 
$|g_{ij}|$ are much smaller than the
$\epsilon_i$; and (ii) certain single-qubit operations such as $\pi$ pulses 
are not used (however, $2 \pi$ pulses are permitted and turn out to be 
extremely useful, and $\pi$ pulses can be used to prepare SES states from the
ground state $|00 \cdots 0\rangle$). 

The advantage of working in the SES can be understood from the following 
expression \cite{PritchettPre10,matrixelementsNote} for the
SES matrix elements of model (\ref{tcsq model}),
\begin{widetext}
\begin{equation}
\big( m \big| H_{\rm qc} \big|m' \big) = \bigg[ \epsilon_m  
- 2J^{zz} \big( \sum_{k<m} g_{km} + \sum_{k>m} g_{mk} \big) 
+ J^{zz} \big(\sum_{i<j} g_{ij}\big)  \bigg] \delta_{mm'} 
+ \bigg[J^{xx} + J^{yy} - i (J^{xy}-J^{yx}) \bigg] g_{mm'}.
\label{ses matrix elements}
\end{equation}
\end{widetext}
Therefore, we have a high degree of control over the part of the Hamiltonian 
in the SES.
For example, in the simple case of an array of qubits coupled with a tunable 
$\sigma^{x}_i \otimes \sigma^{x}_j$ exchange interaction, 
\begin{equation}
\big( m \big| H_{\rm qc} \big|m' \big) =  \epsilon_m   \delta_{mm'}  + g_{mm'},
\label{xx case matrix elements}
\end{equation}
so the diagonal and off-diagonal elements are directly and independently 
controlled by the qubit energies and coupling strengths, respectively. 
Because of this high degree of controllability, $n$-dimensional unitary 
operations can be carried out in a single step, bypassing the need to 
decompose into elementary gates. 
This property also enables the direct quantum simulation of real but otherwise 
arbitrary time-dependent Hamiltonians, allowing a polynomial quantum speedup 
with sub-threshold-fidelity qubits \cite{PritchettPre10}.

\begin{figure}
\includegraphics[width=6.0cm]{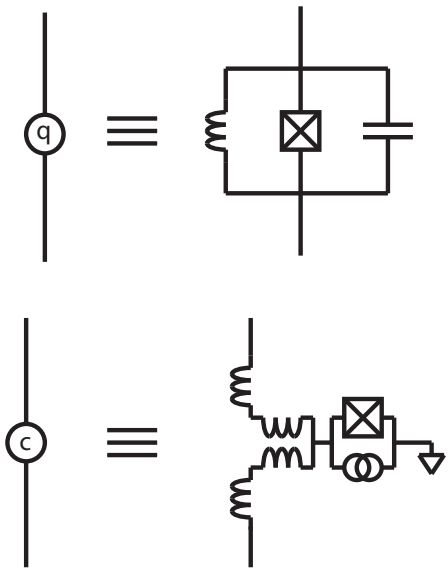} 
\caption{
Definitions of two-terminal phase qubit circuit element ``q" and coupler circuit
element``c", expressed 
in terms of their conventional circuit diagrams involving inductors, capacitors, 
current biases, and 
Josephson junctions (crossed boxes) \cite{PintoPRB10,BialczakPRL11}.}
\label{coupler figure}
\end{figure}

\begin{figure}
\includegraphics[width=7.0cm]{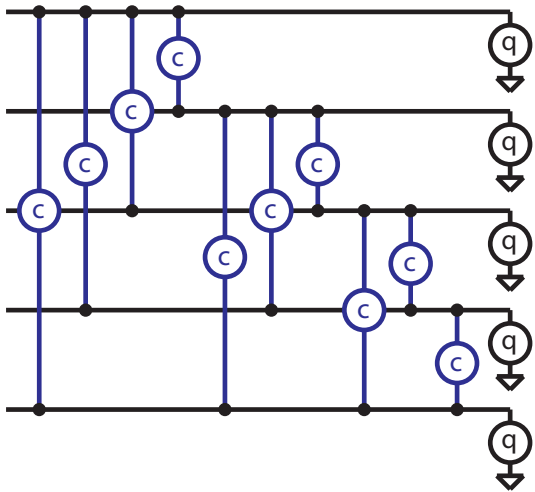} 
\caption{
(color online) Design for a fully connected network of $n$ Josephson 
phase qubits. Each 
qubit ``q"  is suplied with a wire to which $n-1$ coupler leads are attached. 
The case of $n=5$
is shown. The proposed design minimizes crossovers.}
\label{layout figure}
\end{figure}

A quantum computer operating in the SES mode described here requires every 
qubit to be tunably coupled to
the others, as implied by model (\ref{tcsq model}). This requires $n(n-1)/2$ 
coupling wires and associated circuits.
Consider, for example, the recently demonstrated tunable inductive coupler 
for superconducting phase 
qubits \cite{PintoPRB10,BialczakPRL11} (other tunable couplers for superconducting 
qubits have also been 
successfully demonstrated \cite{vanderPloegPRL07,NiskanenSci07,YamamotoPRB08,AllmanPRL10}). 
The circuit diagrams for a single phase qubit ``q" and single coupler 
``c" are illustrated in Fig.~\ref{coupler figure}, 
where the crossed boxes represent Josephson junctions. In terms of these 
elements, a possible layout for a fully connected array is shown in 
Fig.~\ref{layout figure} for the case
of $n=5$. 

The SES computation method is not scalable, because it requires a qubit 
for every Hilbert space
dimension used. However the advantage is that high-dimensional unitary 
operations can be carried
out in a single step without the need to decompose the operation into 
elementary gates. This 
property allows processors of even modest sizes to perform quantum computations
that would otherwise (using a gate-based approach) require thousands of elementary
gates and therefore error-corrected qubits. In the remainder of this paper, 
we illustrate the SES computation 
method by applying it to Grover's search algorithm \cite{GroverPRL97}.

\section{application to Grover's search algorithm}

Here we consider the search for a single marked item in a database of size $n$, 
where $n$ is the
number of qubits in the SES processor (\ref{tcsq model}). Within the SES, 
Grover's algorithm \cite{GroverPRL97} is represented by the following control sequence,
\BEq
\big({W}{\rm O}_{m'}\big)^{n_{\rm Grover}}U_{\rm unif}
|1) \approx |m'),
\EEq
where $U_{\rm unif}$ generates the uniform superposition of all SES states $|m)$, 
$m=1,2,\dots,n$. Here ${\rm O}_{m'}$ is the oracle corresponding to the marked 
state $|m')$, $W$ is Grover's inversion operator, and $n_{\rm Grover} \approx (\pi/4)\sqrt{n}$
is the number of iteration steps of the algorithm. In the following sections we 
show how each of these unitary operators can be generated in a {\it single} step.

\subsection{Preparation of the uniform state}

The uniform superposition state,
\BEq
|\psi_{\rm unif}) \equiv \frac{1}{\sqrt{n}}\sum_{m=1}^n |m) =
\frac{1}{\sqrt{n}}
\begin{bmatrix}
1&1&1&1&\dots 
\end{bmatrix}^{\rm T},
\EEq
can be generated from the state 
$|1) =
\begin{bmatrix}
1 & 0 & 0 & 0 & \dots 
\end{bmatrix}^{\rm T}$,
where $^{\rm T}$ stands for transposition, through
\BEqA
\label{eq:psi_unif}
|\psi_{\rm unif})=ie^{i\a_{\rm unif}} e^{-iH_{\rm unif}t_{\rm unif}}|1),
\label{uniform state identity}
\EEqA 
with
\begin{equation}
\a_{\rm unif} = \pi/(2\sqrt{n})
\end{equation}
and
\begin{equation}
t_{\rm unif} = \pi/(2g\sqrt{n}),
\end{equation}
using the SES Hamiltonian 
\BEqA
  H_{\rm unif} &=& 
g \begin{bmatrix}
		2&  1 & 1 & 1&\dots \cr
    1&  0 & 0 & 0&\dots \cr
    1&  0 & 0 & 0&\dots \cr
    1&  0 & 0 & 0&\dots \cr
    \vdots&\vdots &\vdots &\vdots &\ddots  \cr
 \end{bmatrix}.
 \EEqA
This can be seen by noticing that the spectrum of $H_{\rm unif}$ and the 
transformation $S$ (whose columns are shown unnormalized here for notational 
simplicity) that diagonalizes $H_{\rm unif}$ via 
$H_{\rm unif}^{\rm diag}=S^{\dagger} H_{\rm unif}S$ are given by
 \BEq
 E = 1\mp\sqrt{N}, 0, \dots, 0,
 \EEq
and
\BEq
S = \begin{pmatrix}
		1-\sqrt{N} &1+\sqrt{N}& 0&0&0&\dots &0\cr
           1& 1 &-1 &-1&-1&\dots &-1\cr
           1 &1& 1 &0&0&\dots &0\cr
            1&1&0&1&0&\dots &0 \cr
            1&1&0&0&1&\dots &0\cr
            \vdots&\vdots &\vdots &\vdots &\vdots &\ddots &\vdots \cr
            1&1&0&0&0&\dots&1
            \end{pmatrix},
 \EEq
 respectively. Direct exponentiation then immediately leads to Eq. (\ref{eq:psi_unif}).
 The initial state $|1)$ required in (\ref{uniform state identity}) is easily 
 prepared from the 
 ground state via a $\pi$ pulse.
 
\subsection{Single-step oracle and $W$ operators}
 
The oracle (corresponding to the marked state with $m'=3$, for instance), is
\BEqA
  {\rm O} &=& 1 - 2|m')(m'| = 
 \begin{bmatrix}
		1&  0 & 0 & 0&\dots \cr
    0& 1 & 0 & 0&\dots \cr
    0&  0 &-1 & 0&\dots \cr
    0&  0 & 0 &1 &\dots \cr
    \vdots&\vdots &\vdots &\vdots &\ddots  \cr
 \end{bmatrix},
 \EEqA
which can be generated via 
\BEq
W = e^{-i H_{\rm O} t_{\rm O}}, \quad t_{\rm O} = \pi/ \delta \epsilon,
\EEq 
using the SES Hamiltonian
\BEqA
  H_{\rm O} &=& 
 \begin{bmatrix}
		0&  0 & 0 & 0&\dots \cr
    0&  0 & 0 & 0&\dots \cr
    0&  0 &-\delta \epsilon & 0&\dots \cr
    0&  0 & 0 & 0 &\dots \cr
    \vdots&\vdots &\vdots &\vdots &\ddots  \cr
 \end{bmatrix}.
 \EEqA
 $\delta \epsilon$ is the detuning of qubit $m'$ from the remaining qubits, 
 all having common frequency $\epsilon$.
 This operation is simply a $2\pi$ rotation on the qubit associated with the 
 marked state $m'$. (Although we have
 implemented it as a $z$ rotation, an $x$ or $y$ rotation would work equally 
 well.)
 Notice that in the limit 
$g \ll \epsilon$, which guarantees good isolation of the SES, the oracle can 
be made arbitrarily fast if we choose sufficiently large 
detuning $\delta \epsilon$.

Finally, the $W$ operator,
\BEqA
  {\rm W} &=& 2|\psi_{\rm unif})(\psi_{\rm unif}| - 1
  \nonumber \\
  &=& 
  \frac{1}{n}\begin{bmatrix}
		2-n&  2 & 2 & 2&\dots \cr
    2& 2-n & 2 & 2&\dots \cr
    2&  2 &2-n & 2&\dots \cr
    2&  2 & 2 &2-n &\dots \cr
    \vdots&\vdots &\vdots &\vdots &\ddots  \cr
 \end{bmatrix},
\EEqA
can be generated via 
\BEqA
{\rm W} = e^{-i \a_{\rm W}}e^{-i H_{\rm W} t_{\rm W}},
\nonumber \\
\a_{\rm W} = (1-n)\pi/n, \quad 
t_{\rm W} = \pi/(ng),
\EEqA
with
\BEqA
  H_{\rm W} &=& 
 g \begin{bmatrix}
		0&  1 & 1 & 1&\dots \cr
    1&  0 & 1 & 1&\dots \cr
    1&  1 & 0 & 1&\dots \cr
    1&  1 & 1 & 0 &\dots \cr
    \vdots&\vdots &\vdots &\vdots &\ddots  \cr
 \end{bmatrix}.
 \EEqA
 
\subsection{Comparison with gate-based computation}

Here we compare the SES and conventional gate-based approaches for 
a $n=256$ item search.
The search requires 12 iterations.

In the SES case, we choose a weak, $g/2\pi = 1.25$ MHz coupling, which
guarantees that the $W$ operation is at least 1.5 ns long. 
We also choose a strong, $\delta \epsilon/2\pi = 100$ MHz 
qubit detuning to guarantee that the oracle is sufficiently fast. 
Then $t_{\rm unif} = 12.5$ ns, 
$t_{\rm O} = 5$ ns, 
$t_{\rm W} = 1.56$ ns, and the total duration of the algorithm
is about 100 ns, which is experimentally practical with current 
superconducting architectures.

The SES estimate should be contrasted with the conventional approach, 
which requires an input register of 8 qubits, an output register of 1 
qubit, plus 7 additional ancilla qubits (to make the multiply controlled 
gates more efficient). 
The corresponding search oracle involves an 8-fold CNOT gate, 
${\rm C}^8{\rm NOT}$, each of which can be made out of 85 two-qubit 
CNOTs (using 7 ancillas) \cite{Neilsen2000}. The $W$ operator involves 
a 7-fold controlled-Z gate, ${\rm C}^7{\rm Z}$, which uses 73 standard 
CNOTs (plus 6 ancillas). Thus, a conventional 256-item gate-based Grover 
search requires 158 CNOT gates per search step. 
The full algorithm then contains nearly 2000 CNOT gates, plus local 
rotations.

\section{discussion}

We have described an approach to superconducting quantum computation in which
the computation is carried out in the single-excitation subspace of the 
full Hilbert space. Relative 
to the standard gate-based approach, the SES method requires exponentially 
more qubits and is therefore 
nonscalable. The hardware requirements are also highly demanding: The fully 
connected array of $n$
qubits  requires $n(n-1)/2$ coupling wires and tuning circuits. 
But the SES approach is much 
more time efficient, permitting far larger computations than currently 
possible using today's  
sub-theshold-fidelity qubits in the standard way. This was illustrated 
above for Grover's search
algorithm. We would expect a similar result for Shor's algorithm, but 
note that practical factoring applications
would require impossibly large processor sizes. General purpose 
time-dependent quantum simulation 
can also be carried out in the SES, allowing an $n^3$ polynomial 
quantum speedup \cite{PritchettPre10}. 
However, when a large scale error-corrected quantum computer 
eventually becomes available, 
the gate-based approach will perform better than the SES method.

\newpage

\begin{acknowledgments}

This work was supported by the National Science Foundation under CDI grant DMR-1029764.
It is a pleasure to thank
Joydip Ghosh, John Martinis, Emily Pritchett, Andrew Sornborger  and Phillip Stancil
for useful discussions.

\end{acknowledgments}

\bibliography{/Users/mgeller/Desktop/publications/bibliographies/MRGpre,
/Users/mgeller/Desktop/publications/bibliographies/MRGbooks,
/Users/mgeller/Desktop/publications/bibliographies/MRGgroup,
/Users/mgeller/Desktop/publications/bibliographies/MRGqc-josephson,
/Users/mgeller/Desktop/publications/bibliographies/MRGqc-architectures,
/Users/mgeller/Desktop/publications/bibliographies/MRGqc-general,
/Users/mgeller/Desktop/publications/bibliographies/MRGqc-algorithms,
/Users/mgeller/Desktop/publications/bibliographies/MRGqc-simulation,endnotes}

\begin{thebibliography}{15}
\bibitem[{not()}]{notationNote}
\bibinfo{note}{The $\sigma^\mu$ (with $\mu=x,y,x$) are Pauli matrices, and $$c
  \equiv \begin{pmatrix} 0 & 1 \cr 0 & 0 \end{pmatrix}.$$}.
\bibitem[{mat()}]{matrixelementsNote}
\bibinfo{note}{The expression (\ref{ses matrix elements}) differs from that of
  Ref.~\cite{PritchettPre10} because the tunable inductive coupler considered
  there generates single-qubit as well as two-qubit interactions, a technical
  complication that is not important for the present analysis. Also note that
  the term $J^{zz} \big(\sum_{i<j} g_{ij}\big) \delta_{mm'}$ is an energy shift
  and can be dropped.}
\end{thebibliography}

\end{document}